\documentclass{article}
\usepackage{amsmath}
\usepackage{hhline}
\usepackage{graphicx}
\usepackage[T2A]{fontenc}
\usepackage[cp1251]{inputenc}
\usepackage[russian]{babel}

\begin{document}          \parindent=13pt

\def\ds{\displaystyle}    \def\ss{\scriptstyle}
\def\hh{\hskip 1pt}       \def\hs{\hskip 2pt}   \def\h{\hskip 0.2mm}
\def\pr{\prime}
\newcommand{\blR}{\hbox{\ss\bf R}}
\newcommand{\fbr}{\hbox{\footnotesize\bf r}}
\newcommand{\fbk}{\hbox{\footnotesize\bf k}}
\newcommand{\Tr}{\mathop{\rm T\h r}\nolimits}
\newcommand{\fbR}{\hbox{\footnotesize\bf R}}

\setcounter{page}{1} \makeatletter\renewcommand{\@oddhead} { \hfil
---\hs\hs\hs\thepage\hs\hs\hs ---\hskip 2mm\hfil }
\makeatletter\renewcommand{\@oddfoot}{}

\hskip-13pt {\Large\bf Quantum master equation for a system of
identical particles } \vskip 2mm

\hskip-13pt {\Large\bf and the anisotropic distribution
 of the interacting electrons } \vskip 5mm

\hskip-13pt Boris V. Bondarev \vskip 5mm

\hskip-13pt{\it Moscow Aviation Institude, Volokolamskoye Shosse 4,
125871 Moscow, Russia}\vskip 7mm \hrule\vskip 5mm

\hskip -13pt A R T I C L E  \hskip 10mm I N F O \hskip 10mm A B S T
R A C T \vskip 0mm
\begin{picture}(0,0)\put(-13,0){\line(1,0){310}}\end{picture}
\begin{picture}(160,0)\put(145,0){\line(1,0){295}}\end{picture}
\vskip 2mm\hskip -13pt {\it Article history:} \vskip 5mm
\begin{picture}(0,0)\put(-13,0){\line(1,0){150}}\end{picture}
\vskip 2mm\hskip -13pt {\it Keywords:}

\vskip 0mm\hskip -13pt Density matrix for $N$

\vskip 0mm\hskip -13pt particles

\vskip 0mm\hskip -13pt Equation with dissipative

\vskip 0mm\hskip -13pt terms

\vskip 0mm\hskip -13pt Hierarchy of density matrices

\vskip 0mm\hskip -13pt Equation for single-particle

\vskip 0mm\hskip -13pt density matrix

\vskip 0mm\hskip -13pt Mean-field approximation

\vskip 0mm\hskip -13pt Anisotropy of distribution

\vskip 0mm\hskip -13pt Interacting electrons

\vskip -58mm\hskip 46mm \qquad\parbox{10.3cm}{\hskip 13pt We
consider an open quantum many-particle system in which there are
dissipative processes. The evolution of this system is described
by a kinetic equation for the density matrix. From the equation
describing a random Markov process in this system, we obtain an
equation for the single-particle statistical operator. This
equation describes the evolution of a system of identical
particles in a mean-field approximation. The equation for
interacting particles in thermodynamic equilibrium was obtained.
The distribution function of a system of interacting electrons in
metals has multivalence in a certain region of wave vectors. Among
many solutions one is isotropic. Other solutions have the
anisotropy of the electron distribution over the wave vectors. The
anisotropy arises as a result of repulsion and attraction between
electrons. }

\vskip 5mm\hskip 97mm 2012 Elsevier B.V. All rights reserved

\vskip 5mm
\begin{picture}(0,0)\put(-13,0){\line(1,0){150}}\end{picture}
\begin{picture}(160,0)\put(145,0){\line(1,0){295}}\end{picture}

\vskip 5mm \hskip-13pt{\bf 1. Introduction} \vskip 3mm

The most general statistical description of any system in quantum
mechanics is performed by using the statistical operator [1]. If
the system state changes with time, then the statistical operator
is a function of time:
$$ \hat\varrho=\hat\varrho\hh(\hh t\hh )\hs . \eqno (1.1) $$ \par
When the system takes a random Markov process, this function can
be found from the kinetic equation of the form [2 – 14] $$
i\hs\hbar\hskip 1.1mm\dot{\hskip -0.5mm\hat\varrho}=
\bigl[\hh\widehat H,\hh\hat\varrho\hs\bigr]+ i\hs\hbar\hs\hat
d\hs\hat\varrho\hs , \eqno (1.2) $$ which is called the Liouville
- von Neumann equation. In this equation, $\widehat H$ is the
Hamiltonian of the system; $\hat d$ is the so-called dissipative
operator describing dissipative processes in the system. The
statistical operator should be normalized, self-adjoint and
positive. Any solution of (1.2) that has a physical significance
and has these properties at the initial moment of time $t_0$
should have them for all $t>t_0$. So the equation (1.2) should
have the form
$$ i\hs\hbar\hskip 1.1mm\dot{\hskip -0.5mm\hat\varrho}=
\bigl[\hh\widehat H,\hh\hat\varrho\hs\bigr]+ i\hs\hbar\hs
\sum\limits_{k,\hh l}\hs C_{k\hh l}\hs\Bigl(\hh \bigl[\hs\hat
a_k\hs\hat\varrho\hh ,\hs\hat a_l^{\hh +}\bigl]+ \bigl[\hs\hat
a_k\hh ,\hs\hat\varrho\hs\hh\hat a_l^{\hh +}\bigl]\hs\Bigr)\hs .
\eqno (1.3) $$ This equation is called the Lindblad equation [6].
Here $a_k$ is an operator, $k = 1, 2, ...$ – the number of the
operator, $C_{k\hh l}$ - constants.
\par Let us write the equation
(1.3) for the quantum harmonic oscillator: $$ i\hs\hbar\hskip
1.1mm\dot{\hskip -0.5mm\hat\varrho}= \bigl[\hh\widehat
H,\hh\hat\varrho\hs\bigr] +\frac{1}{2}\hs i\hs\hbar\hs
A\hs\Bigl(\hh \bigl[\hs\hat a\hs\hat\varrho\hh ,\hs\hat a^{\hh
+}\bigl]+ \bigl[\hs\hat a\hh ,\hs\hat\varrho\hs\hh\hat a^{\hh
+}\bigl]\hs\Bigr) +\frac{1}{2}\hs i\hs\hbar\hs B\hs\Bigl(\hh
\bigl[\hs\hat a^{\hh +}\hs\hat\varrho\hh ,\hs\hat a\bigl]+
\bigl[\hs\hat a^{\hh +} ,\hs\hat\varrho\hs\hh\hat
a\bigl]\hs\Bigr)\hs , \eqno (1.4) $$ where $$ \widehat
H=\hbar\hs\omega\Biggl(\hat a^{\hh +}\hs\hat a+\frac{1}{2}
\Biggr)\hs , \eqno (1.5) $$ $A$ and $B$ are constants. The
operator $\hat a$ has the form $$ \hat a=\frac{1}{\sqrt{\hh
2\hs\hbar\hs\omega\hh }}\hs\Biggl( \hh\frac{i\hs\hat p }{\sqrt{\hh
m}}+\sqrt{\hh\kappa}\hs\hat x\Biggr)\hs. \eqno (1.6) $$ Equation
(1.4) surprisingly accurately describes the change of the state of
the quantum harmonic oscillator with time, provided that $$
\frac{A}{\hh B\hh}=e^{\hh\beta\hh\hbar\hh\omega\hh}\hs , \eqno
(1.7)$$ where $\beta=1/{\hh k\hs T\hh}$ is the inverse
temperature.

\vskip 5mm \hskip-13pt{\bf 2. The hierarchy of statistical
operators} \vskip 3mm Consider the system consisting of $N$
identical particles. The statistical operator $\hat\varrho$
describing the state of this system can be written as follows:
$$ \hat\varrho =\hat\varrho\hh(\hh 1,\hs 2,...,\hh N\hh )\hs .
\eqno (2.1) $$ Here the numbers in parentheses denote the indices
of the variables which are affected by this operator. The
statistical operator like any other operator describing the state
of the system of identical particles  must be symmetric: $$
\hat\varrho\hh (\hh ...,\hs i,...,\hs j,... \hh )=
   \hat\varrho\hh (\hh ...,\hs j,...,\hs i,... \hh )\hs . \eqno (2.2) $$

\par Let us assume the following normalization condition for the statistical operator:
$$ \Tr_{\hh 1\hh ...\hh N}^{\phantom{\underline J}}
\hat\varrho\hh (\hh 1,...,\hh N)=N!\hs . \eqno (2.3) $$ \par If
the system state changes over time, the statistical operator is a
function of time: $t$: $ \hat\varrho =\hat\varrho\hh (t)\hh . $
When we can be sure that the system takes a random Markov process,
this function can be found from the equation $$ i\hs\hbar\hskip
1.1mm\dot{\hskip -0.5mm\hat\varrho}= \bigl[\hh\widehat{\cal
H},\hh\hat\varrho\hs\bigr]+ i\hs\hbar\hs\hat d\hs\hat\varrho\hs .
\eqno (2.4) $$ The Hamiltonian operator in this equation can be
written as the sum $$ \widehat {\cal H}\hh (\hh 1,...,\hh N\hh
)=\sum\limits_{i=1}^N\widehat H(i\hh )+ \frac{1}{\hs
2\hs}\hs\sum\limits_{i=1}^N \sum\limits_{\ss j=1\atop\ss j\neq
i}^N \widehat H(i,\hh j\hh)\hs , \eqno (2.5) $$ where $\widehat
H(i\hh )$ is a single-particle Hamiltonian, i.e. the energy
operator of a single particle without considering its interaction
with other particles; $\widehat H(i,\hh j\hh)$ is the operator of
the interaction of two particles. A single-particle Hamiltonian
may contain dissipative terms. A two-particle Hamiltonian should
be symmetric:
$$ \widehat H(i,\hh j\hh)=\widehat H(\hh j,\hh i\hh)\hs . \eqno
(2.6) $$ Under this condition, the Hamiltonian (2.5) will also be
symmetric.
\par
Suppose that the action of the dissipative operator $\hat d$ on
the statistical operator $\hat\varrho$ leads to the expression $$
\hat d\hs\hat\varrho =\frac{1}{\hh 2\hh}\hs\Bigl(\hh \bigl[\hs\hat
a\hs\hat\varrho\hh ,\hs\hat a^{\hh +}\bigl]+ \bigl[\hs\hat a\hh
,\hs\hat\varrho\hs\hh\hat a^{\hh +}\bigl]\hs\Bigr)\hs . \eqno
(2.7)
$$ Here the indices $k$ and $l$ near the operator $\hat
a$ are omitted .
\par
Suppose that the dissipative processes occurring in the system are
caused by the stochastic interaction of particles with a heat
reservoir. Moreover, each particle interacts with it independently
of other particles. In this case the many-particle operator $\hat
a$ in the expression (2.7) can be written as the sum $$ \hat a\hh
(\hh 1,...,\hh N\hh )= \sum\limits_{i=1}^N\hat a\hh(i\hh )\hs ,
\eqno (2.8) $$ where the operator $\hat a\hh(i\hh )$ characterizes
the effect of the heat reservoir on one of the particles.
\par
Substitution of the sum (2.8) in the formula (2.7) gives $$ \hat
d\hs\hat\varrho = \frac{1}{\hh 2\hh}\hs\sum\limits_{i=1}^N\hh
\sum\limits_{j=1}^N\Bigl(\hh\bigl[\hs\hat a\hh (i\hh )\hs
\hat\varrho\hh ,\hs\hat a^{\hh +}(\hh j\hh)\ hh\bigl]\hs +
\bigl[\hs\hat a\hh (i\hh)\hh ,\hs\hat\varrho\hs\hh \hat a^{\hh
+}(\hh j\hh)\hh\bigl]\hs\Bigr)\hs . \eqno (2.9) $$
\par
It is impossible to solve the equation (2.4) with such a complex
right-hand side. Therefore the practical interest is carried by
the single-particle kinetic equation for the statistical operator
$\hat\varrho\hh (1)$, which is defined by the relation $$
\hat\varrho\hh (1)=\frac{1}{\hh (\hh N-1\hh)!\hh}\hs \Tr_{\hh 2\hh
...\hh N}^{\phantom{\underline J}} \hat\varrho\hh (\hh 1,\hs
2,...,\hh N)\hs . \eqno (2.10) $$ This definition and condition
(2.3) implies the normalization condition $$ \Tr_{\hh
1}^{\phantom{\underline J}} \hat\varrho\hh (\hh 1)=N\hs . \eqno
(2.11) $$
\par
We define the two-particle statistical operator
$\hat\varrho\hh (1,\hh2)$ as $$ \hat\varrho\hh (1,\hh
2)=\frac{1}{\hh (\hh N-2\hh)!\hh}\hs \Tr_{\hh 3\hh ...\hh
N}^{\phantom{\underline J}} \hat\varrho\hh (\hh 1,\hs 2,...,\hh
N)\hs . \eqno (2.12) $$ This operator satisfies the normalization
condition $$ \Tr_{\hh 1\hh 2}^{\phantom{\underline J}}
\hat\varrho\hh (\hh 1,\hh 2)=N\hs (\hh N-1\hh)\hs . \eqno (2.13)
$$

\vskip 7mm \hskip-13pt{\bf 3. The quantum kinetic equation for
single-particle density matrix } \vskip 3mm To obtain the equation
for the single-particle operator $\hat\varrho\hh (1)$, let us
apply the operation of coagulation $\Tr_{\hh 2\hh ...\hh N}$ to
both sides of the equation (2.4). As $$ \Tr_{\hh
i}^{\phantom{\underline J}}\hh\bigl[\hh\widehat H(i\hh ),\hs
\hat\varrho\hh (\hh 1,\hh ...,\hh N)\hh\bigr]\equiv 0 $$ by
definition (2.10), we obtain $$ \Tr_{\hh 2\hh ...\hh
N}^{\phantom{\underline J}}\hs \sum_{i=1}^N\hh\bigl[\hh \widehat
H(i\hh ),\hs\hat\varrho\hh (\hh...,\hs i,...)\hh\bigr]= (\hh
N-1\hh)!\hs\bigl[\hh\widehat H(1),\hs\hat\varrho\hh
(1)\hh\bigr]\hs . $$
\par
Due to the identity
$$ \Tr_{\hh i\hh j}^{\phantom{\underline J}}\hh \bigl[\hh\widehat
H(i,\hh j\hh), \hs\hat\varrho\hh (\hh 1,\hh ...,\hh
N)\hh\bigr]\equiv 0\hs , $$ taking into account the property (2.6)
and the definition (2.12) we have $$ \Tr_{\hh 2\hh ...\hh
N}^{\phantom{\underline J}}\hs \frac{1}{\hs 2\hs}\hs
\sum\limits_{i=1}^N\sum\limits_{\ss j=1\atop\ss j\neq i}^N\hh
\bigl[\hh\widehat H(i,\hh j\hh), \hs\hat\varrho\hh (\hh 1,\hh
...,\hh N)\hh\bigr]\hs = $$ $$ =\hs\Tr_{\hh 2\hh ...\hh
N}^{\phantom{\underline J}}\hs\Biggr(\hh
\sum\limits_{j=2}^N\hs\bigl[\hh\widehat H(1,\hh j\hh),
\hs\hat\varrho\hh (\hh 1,\hh ...,\hh N)\hh\bigr]\hs + \frac{1}{\hs
2\hs}\hs \sum\limits_{i=2}^N\sum\limits_{\ss j=2\atop\ss j\neq
i}^N\hh \bigl[\hh\widehat H(i,\hh j\hh), \hs\hat\varrho\hh (\hh
1,\hh ...,\hh N)\hh\bigr]\Biggl)\hs = $$ $$ =\hs (\hh
N-1\hh)!\hs\Tr_{\hh 2}^{\phantom{\underline J}}\hs
\bigl[\hh\widehat H(1,\hh 2\hh),\hs\hat\varrho\hh (1,\hh
2\hh)\hh\bigr]\hs . $$ Similarly $$ \Tr_{\hh 2\hh ...\hh
N}^{\phantom{\underline J}}\hs\hat d\hs \hat\varrho\hs =
\hs\frac{1}{\hh 2\hh}\hs \Tr_{\hh 2\hh ...\hh
N}^{\phantom{\underline J}}\hs\Biggl( \bigl[\hs\hat a\hh (1\hh
)\hs\hat\varrho\hh ,\hs \hat a^{\hh +}(\hh
1\hh)\hh\bigl]+\bigl[\hs\hat a\hh (1\hh)\hh ,\hs
\hat\varrho\hs\hh\hat a^{\hh +}(\hh 1\hh)\hh\bigl]\hs + $$ $$ +\hs
\sum\limits_{i=2}^N\hh\Bigl(\hh \bigl[\hs\hat a\hh (i\hh
)\hs\hat\varrho\hh ,\hs \hat a^{\hh +}(\hh
1\hh)\hh\bigl]+\bigl[\hs\hat a\hh (i\hh)\hh ,\hs
\hat\varrho\hs\hh\hat a^{\hh +}(\hh 1\hh)\hh\bigl]\hs + \hs
\bigl[\hs\hat a\hh (1\hh )\hs\hat\varrho\hh ,\hs \hat a^{\hh
+}(\hh i\hh)\hh\bigl]+\bigl[\hs\hat a\hh (1\hh)\hh ,\hs
\hat\varrho\hs\hh\hat a^{\hh +}(\hh i\hh)\hh\bigl]\hh\Bigr)\hs +
$$
$$ +\hs\sum\limits_{i=2}^N\hh
\sum\limits_{j=2}^N\Bigl(\hh\bigl[\hs\hat a\hh (i\hh )\hs
\hat\varrho\hh ,\hs\hat a^{\hh +}(\hh j\hh)\hh\bigl] + \bigl[\hs\hat
a\hh (i\hh)\hh ,\hs\hat\varrho\hs\hh \hat a^{\hh +}(\hh
j\hh)\hh\bigl]\hs\Bigr)\Biggr)\hs = $$ $$ =\hh\frac{1}{\hh 2\hh}\hh
(\hh N-1\hh)!\hh\Bigl(\hh \bigl[\hh\hat a\hh (1\hh
)\hs\hat\varrho\hh(1),\hs \hat a^{\hh +}(\hh
1\hh)\hh\bigl]+\bigl[\hs\hat a\hh (1\hh),\hs
\hat\varrho\hh(1)\hs\hh\hat a^{\hh +}(\hh 1\hh)\hh\bigl]\hh +
\hs\bigl[\hs\Tr_{\hh 2}^{\phantom{\underline J}}\hh \hat a\hh
(2\hh)\hs\hat\varrho\hh(1,\hh 2),\hs \hat a^{\hh +} (\hh
1\hh)\hh\bigl]\hs + $$ $$ +\hs \bigl[\hs\hat a\hh (1\hh),\hs
\Tr_{\hh 2}^{\phantom{\underline J}}\hh \hat\varrho\hh(1,\hh
2)\hs\hh\hat a^{\hh +}(\hh 2\hh)\hh\bigl]\hh\Bigr)\hs , $$ as $$
\Tr_{\hh i}^{\phantom{\underline J}}\hh\bigl[\hs\hat a\hh (i\hh)\hh
,\hs \hat\varrho\hs\hh\hat a^{\hh +}(\hh j\hh)\hh\bigl]\hh\equiv\hh
0\hs , \hskip 15mm\Tr_{\hh j}^{\phantom{\underline
J}}\hh\bigl[\hs\hat a\hh (i\hh )\hs \hat\varrho\hh ,\hs\hat a^{\hh
+}(\hh j\hh)\hh\bigl]\hh\equiv\hh 0\hs . $$
Bringing together these
expressions, we obtain the equation $$ i\hs\hbar\hskip
1.1mm\dot{\hskip -0.5mm\hat\varrho}\hh (1)= \bigl[\hh\widehat
H(1),\hh\hat\varrho\hh(1)\hs\bigr]+ \Tr_{\hh 2}^{\phantom{\underline
J}}\hs \bigl[\hh\widehat H(1,\hh 2\hh), \hs\hat\varrho\hh (1,\hh
2\hh)\hh\bigr]\hs + $$ $$ +\hs\frac{1}{\hh 2\hh}\hs
i\hs\hbar\hh\Bigl(\hh \bigl[\hh\hat a\hh (1\hh
)\hs\hat\varrho\hh(1),\hs \hat a^{\hh +}(\hh
1\hh)\hh\bigl]+\bigl[\hs\hat a\hh (1\hh),\hs
\hat\varrho\hh(1)\hs\hh\hat a^{\hh +}(\hh 1\hh)\hh\bigl]\hh +
\hs\bigl[\hs\Tr_{\hh 2}^{\phantom{\underline J}}\hh \hat a\hh (2\hh
)\hs\hat\varrho\hh(1,\hh 2),\hs \hat a^{\hh +}(\hh 1\hh)\hh\bigl]\hs
+ $$ $$ +\hs \bigl[\hs\hat a\hh (1\hh),\hs\Tr_{\hh
2}^{\phantom{\underline J}}\hh \hat\varrho\hh(1,\hh 2)\hs\hh\hat
a^{\hh +}(\hh 2\hh)\hh\bigl]\hh\Bigr)\hh . \eqno (3.1) $$
\par
We denote the matrix elements $\hat\varrho\hh(1)$,
$\hat\varrho\hh(1,\hh 2)$, $\widehat H(1)$, $\widehat H(1,\hh 2)$
and $\hat a(1)$ in an $\alpha$-representation as follows: $$
\varrho_{\alpha_1^{\phantom{\pr}}\alpha_1^\pr}= \varrho_{\hh
11^\pr}\hs , \hskip 15mm
\varrho_{\alpha_1^{\phantom{\pr}}\alpha_2^{\phantom{\pr}},
\hh\alpha_1^\pr \alpha_2^\pr}=\varrho_{\hh 12,\hh 1^\pr 2^\pr}\hs
,
$$ $$ H_{\alpha_1^{\phantom{\pr}}\alpha_1^\pr}=H_{11^\pr}\hs ,
\hskip 15mm H_{\alpha_1^{\phantom{\pr}}\alpha_2^{\phantom{\pr}},
\hh\alpha_1^\pr \alpha_2^\pr}=H_{12,\hh 1^\pr 2^\pr}\hs , \hskip
15mm a_{\alpha_1^{\phantom{\pr}}\alpha_1^\pr}=a_{\hh 11^\pr}\hs , $$
and write the equation (3.1) in matrix form as: $$
i\hs\hbar\hs\dot\varrho_{\hh 11^\pr}=\sum\limits_{\alpha_2}\bigl(\hh
H_{12}\hs\varrho_{\hh 21^\pr}-\varrho_{\hh 12}\hs
H_{21^\pr}\hh\bigr)\hs +
\hs\sum\limits_{\alpha_2}\sum\limits_{\alpha_3}\sum\limits_{\alpha_4}
\bigl(\hh H_{12,\hh 34}\hs\varrho_{\hh 34,\hh 1^\pr 2}- \varrho_{\hh
12,\hh 34}\hs H_{34,\hh 1^\pr 2}\hh\bigr)\hs + $$ $$
+\hs\frac{1}{\hh 2\hh}\hs i\hs\hbar\hs\sum\limits_{\alpha_2}
\sum\limits_{\alpha_3}\bigl(\hh 2\hs a_{\hh
12}^{\phantom{+}}\hs\varrho_{\hh 23}^{\phantom{+}}\hs a_{\hh
31^\pr}^+- a_{\hh 12}^+\hs a_{\hh 23}^{\phantom{+}}\hs\varrho_{\hh
31^\pr}^{\phantom{+}} \hs - \hs\varrho_{\hh 12}^{\phantom{+}}\hs
a_{\hh 23}^+\hs a_{\hh 31^\pr}^{\phantom{+}}\hh\bigr)\hs + $$ $$
+\hs\frac{1}{\hh 2\hh}\hs i\hs\hbar\hs\sum\limits_{\alpha_2}
\sum\limits_{\alpha_3}\sum\limits_{\alpha_4}\Bigl(\hh \varrho_{\hh
12,\hh 34}^{\phantom{+}}\hs\bigl(\hh a_{\hh 42}^{\phantom{+}}\hs
a_{\hh 31^\pr}^+ - a_{\hh 42}^+\hs a_{\hh
31^\pr}^{\phantom{+}}\hh\bigr)\hs + \hs\bigl(\hh a_{\hh
12}^{\phantom{+}}\hs a_{\hh 34}^+ - a_{\hh 12}^+\hs a_{\hh
34}^{\phantom{+}}\hh\bigr)\hs \varrho_{\hh 42,\hh
31^\pr}^{\phantom{+}}\hh\Bigr)\hs . \eqno (3.2) $$
\par
The two-particle density matrix describing the fermion system
state must be antisymmetric. This matrix can be approximately
expressed by means of the single-particle density matrix as
follows:
$$ \varrho_{\hh 12,\hh 1^\pr 2^\pr}^{\phantom{+}}= \varrho_{\hh
11^\pr}^{\phantom{+}}\hs\varrho_{\hh 22^\pr}^{\phantom{+}}-
\varrho_{\hh 12^\pr}^{\phantom{+}}\hs\varrho_{\hh
21^\pr}^{\phantom{+}}\hs . \eqno (3.3) $$ Substituting this
expression into the equation (3.2), we obtain the equation for the
single-particle density matrix $$ i\hs\hbar\hs\dot\varrho_{\hh
11^\pr}=\sum\limits_{\alpha_2}\Bigl(\hh \overline
H_{12}\hs\varrho_{\hh 21^\pr}-\varrho_{\hh 12}\hs \overline
H_{21^\pr}\Bigr)\hs + \hs\frac{1}{\hh 2\hh}\hs
i\hs\hbar\hs\sum\limits_{\alpha_2} \sum\limits_{\alpha_3}\bigl(\hh
2\hs a_{\hh 12}^{\phantom{+}}\hs\varrho_{\hh 23}^{\phantom{+}}\hs
a_{\hh 31^\pr}^+- a_{\hh 12}^+\hs a_{\hh
23}^{\phantom{+}}\hs\varrho_{\hh 31^\pr}^{\phantom{+}} \hs -
\hs\varrho_{\hh 12}^{\phantom{+}}\hs a_{\hh 23}^+\hs a_{\hh
31^\pr}^{\phantom{+}}\hh\bigr)\hs + $$ $$ +\hs\frac{1}{\hh
2\hh}\hs i\hs\hbar\hs\sum\limits_{\alpha_2}
\sum\limits_{\alpha_3}\sum\limits_{\alpha_4}\Bigl(\hh \varrho_{\hh
12}^{\phantom{+}}\hs\varrho_{\hh 34}^{\phantom{+}}\hs \bigl(\hh
a_{\hh 43}^{\phantom{+}}\hs a_{\hh 21^\pr}^+ - a_{\hh 43}^+\hs
a_{\hh 21^\pr}^{\phantom{+}}\hs - \hs a_{\hh 23}^{\phantom{+}}\hs
a_{\hh 41^\pr}^+ + a_{\hh 23}^+\hs a_{\hh
41^\pr}^{\phantom{+}}\hh\bigr)+ $$ $$ +\bigl(\hh a_{\hh
12}^{\phantom{+}}\hs a_{\hh 34}^+ - a_{\hh 12}^+\hs a_{\hh
34}^{\phantom{+}}\hs - \hs a_{\hh 14}^{\phantom{+}}\hs a_{\hh
32}^+ + a_{\hh 14}^+\hs a_{\hh 32}^{\phantom{+}}\hh\bigr)\hs
\varrho_{\hh 43}^{\phantom{+}}\hs\varrho_{\hh
21^\pr}^{\phantom{+}} \hh\Bigr)\hs , \eqno (3.4) $$ where $$
\overline
H_{12}=H_{12}+2\hs\sum\limits_{\alpha_3}\sum\limits_{\alpha_4}
H_{13,\hh 24}\hs\varrho_{\hh 43}\hs . \eqno (3.5) $$
\par
Let us write the equation (3.4) in operator form
$$ i\hs\hbar\hskip
1.1mm\dot{\hskip -0.5mm\hat\varrho}= \bigl[\hh\overline{\widehat
H},\hh\hat\varrho\hs\bigr]+ \frac{1}{\hh 2\hh}\hs
i\hs\hbar\hs\Bigl(\hh \bigl[\hs\hat a\hs\hat\varrho\hh ,\hs\hat
a^{\hh +}\bigl]+ \bigl[\hs\hat a\hh ,\hs\hat\varrho\hs\hh\hat
a^{\hh +}\bigl]\hs\Bigr)\hs + $$ $$ +\hs\frac{1}{\hh 2\hh}\hs
i\hs\hbar\hs\Bigl(\hh \bigl[\hs\Tr_{\hh 2}\hh (\hh\hat
a^+\hat\varrho\hh )\hs\hat a- \Tr_{\hh 2}\hh (\hh\hat
a\hs\hat\varrho\hh )\hs\hat a^+ ,\hs\hat\varrho\hs\bigr] \hs + \hs
\bigl[\hh\hat\varrho\hs\hat a^+,\hs\hat\varrho\hs\hat a\hh\bigr]+
\bigl[\hh\hat a^+\hat\varrho\hh,\hs\hat
a\hs\hat\varrho\hh\bigr]\hh\Bigr) \hs , \eqno (3.6) $$ where
$\hat\varrho=\hat\varrho\hh(1)$, $$ \overline{\widehat H}=\widehat
H(1)+2\hs \Tr_{\hh 2}^{\phantom{\underline J}}\hs\bigl[\hh\widehat
H(1,\hh 2),\hs \hat\varrho\hh(2)\hh\bigr] \eqno (3.7) $$ is an
averaged single-particle Hamiltonian. The equation (3.6) can be
given more compact form if we use the notation $$
\overline{\widehat H}^{\hs\pr} =\overline{\widehat H}+
\frac{1}{\hh 2\hh}\hs i\hs\hbar\hs\Bigl(\hh \Tr_{\hh 2}\hh
(\hh\hat a^+\hat\varrho\hh )\hs\hat a- \Tr_{\hh 2}\hh (\hh\hat
a\hs\hat\varrho\hh )\hs\hat a^+\Bigr)\hs . \eqno (3.8)$$ In this
case we have $$ i\hs\hbar\hskip 1.1mm\dot{\hskip
-0.5mm\hat\varrho}=\bigl[\hh \overline{\widehat H}^{\hs\pr}\hskip
-1mm,\hh\hat\varrho\hs\bigr]\hs + \ds\hs\frac{1}{\hh 2\hh}\hs
i\hs\hbar\hs\Bigl(\hh\bigl[\hs\hat a\hs\hat\varrho \hh ,\hs\hat
a^{\hh +}\hh (\hh 1-\hat\varrho\hh )\hh\bigl]+\bigl[\hh (\hh
1-\hat\varrho\hh )\hs\hat a\hh ,\hs\hat\varrho\hs\hh\hat a^{\hh
+}\bigl] \hh\Bigr)\hs . \eqno (3.9)
$$
\par
In general the right sides of equations (3.8) and (3.9) may
contain several terms describing dissipative effects which depend
on different operators $\hat a_{\hh k}$. Taking this into account
we generalize the equation (3.8) and write it as follows: $$
i\hs\hbar\hskip 1.1mm\dot{\hskip -0.5mm\hat\varrho}=\bigl[\hh
\overline{\widehat H}^{\hs\pr}\hskip
-1mm,\hh\hat\varrho\hs\bigr]\hs + i\hs\hbar\hs\sum\limits_{k,\hh
l}\hs C_{k\hh l}\hs\Bigl(\hh\bigl[\hs\hat a_{\hh k}\hs\hat\varrho
\hh ,\hs\hat a^{\hh +}_{\hh l}\hh (\hh 1-\hat\varrho\hh
)\hh\bigl]+\bigl[\hh (\hh 1-\hat\varrho\hh )\hs\hat a_{\hh k}\hh
,\hs\hat\varrho\hs\hh\hat a^{\hh +}_{\hh l}\bigl] \hh\Bigr)\hs ,
\eqno (3.10) $$ where $$ \overline{\widehat H}^{\hs\pr}
=\overline{\widehat H}+ \hs i\hs\hbar\hs \sum\limits_{k,\hh l}\hs
C_{k\hh l}\hs\Bigl(\hh \Tr_{\hh 2}\hh (\hh\hat a^+_{\hh
l}\hat\varrho\hh )\hs\hat a_{\hh k}- \Tr_{\hh 2}\hh (\hh\hat
a_{\hh k}\hs\hat\varrho\hh )\hs\hat a^+_{\hh l}\Bigr)\hs . \eqno
(3.11) $$
\par
We write the equations (3.10) and (3.11) in matrix
form: $$ i\hs\hbar\hs\dot\varrho_{\hh
11^\pr}=\sum\limits_{\alpha_2}\Bigl(\hh \overline
H_{12}^{\hs\pr}\hs\varrho_{\hh 21^\pr}-\varrho_{\hh 12}\hs
\overline H_{21^\pr}^{\hs\pr}\Bigr)\hs + \hs\frac{1}{\hh 2\hh}\hs
i\hs\hbar\hs\sum\limits_{\alpha_2} \sum\limits_{\alpha_3}\bigl(\hh
2\hs \Gamma_{\hh 12,\hh 31^\pr}^{\phantom{+}}\hs\varrho_{\hh
23}^{\phantom{+}}\hs- \Gamma_{\hh 23,\hh
12}^{\phantom{+}}\hs\varrho_{\hh 31^\pr}^{\phantom{+}} \hs -
\hs\Gamma_{\hh 31^\pr,\hh 23}^{\phantom{+}}\hs \varrho_{\hh
12}^{\phantom{+}}\hh\bigr)\hs + $$ $$ +\hs\frac{1}{\hh 2\hh}\hs
i\hs\hbar\hs\sum\limits_{\alpha_2}
\sum\limits_{\alpha_3}\sum\limits_{\alpha_4}\Bigl(\hh \varrho_{\hh
12}^{\phantom{+}}\hs\varrho_{\hh 34}^{\phantom{+}}\hs
\bigl(\hh\Gamma_{\hh 43,\hh 21^\pr}^{\phantom{+}}- \Gamma_{\hh
21^\pr,\hh 43}^{\phantom{+}}\hs - \hs\Gamma_{\hh 23,\hh
41^\pr}^{\phantom{+}}+ \Gamma_{\hh 41^\pr,\hh
23}^{\phantom{+}}\hh\bigr)\hs + $$ $$ +\hs\bigl(\hh\Gamma_{\hh
12,\hh 34}^{\phantom{+}}- \Gamma_{\hh 34,\hh 12}^{\phantom{+}}\hs
- \hs\Gamma_{\hh 14,\hh 32}^{\phantom{+}}+ \Gamma_{\hh 32,\hh
14}^{\phantom{+}}\hh\bigr)\hs \varrho_{\hh
43}^{\phantom{+}}\hs\varrho_{\hh 21^\pr}^{\phantom{+}}
\hh\Bigr)\hs , \eqno (3.12) $$ where $$ \overline H_{12}^{\hs\pr}
=H_{12}+ \sum\limits_{\alpha_3}\hs\sum\limits_{\alpha_4}\hs
\Bigl(\hh 2\hs H_{13,24}\hs+ \frac{1}{2}\hs
i\hs\hbar\hs\bigl(\hh\Gamma_{\hh 12,\hh 43}^{\phantom{+}}-
\Gamma_{\hh 43,\hh 12}^{\phantom{+}}\hh\bigr)\Bigr)\hs\varrho_{\hh
43} \hs , \eqno (3.13) $$ $$ \Gamma_{\hh 12,\hh
31^\pr}^{\phantom{+}}=2\hs\sum\limits_{k,\hh l} C_{k\hh l}\hs
a_{\hh 12,\hh k}\hs a_{\hh 31^\pr,\hh l}^+\hs . \eqno (3.14) $$

\vskip 7mm \hskip-13pt{\bf 4. The principle of detailed balance }

\vskip 3mm Suppose that at some point in time the density matrix
becomes diagonal: $$ \varrho_{11^\pr}=W_1\hs\delta_{\hh 11^\pr}\hs
, \eqno (4.1) $$ where $W_1\equiv W_{\alpha_1}$ is the probability
of the state $\alpha_1$ being filled by a particle. The
substitution of matrix (4.1) into (3.12) leads to the equation $$
\dot W_1=\sum\limits_{\alpha_2}\hh\Bigl(\hh P_{\hh 12}\hs (\hh
1-W_1)\hs W_2-P_{\hh 21}\hs (\hh 1-W_2)\hs W_1\hh\Bigr) \hs ,
\eqno (4.2) $$ where $$ P_{\hh 12}=\Gamma_{\hh 12,\hh
21}^{\phantom{+}}\hs \eqno (4.3) $$ is the  probability of the
transition of a particle  from state $\alpha_2$ to state
$\alpha_1$ during a time unit. The transition probability $P_{\hh
12}$ can always be represented as $$ P_{\hh 12}=P_{\hs 12}^{\hh
(o)}\hs e^{\hh -\hh\frac{1}{2}\hh\beta\hh
(\overline\varepsilon_1-\overline\varepsilon_2)}\hs , \eqno (4.4)
$$ where $$ P_{\hs 12}^{\hh (o)}=P_{\hs 21}^{\hh (o)}\hs , $$
$\overline\varepsilon_i$ – the average energy of a particle in the
state $\alpha_i$, which is an eigenvalue of operator (3.13): $$
\overline\varepsilon_1=\varepsilon_1+ 2\hs\sum\limits_{\alpha_2}
H_{12,\hh 12}\hs W_2\hs , \eqno (4.5) $$ where
$\varepsilon_1=H_{11}$.
\par
When a fermion system comes to the state of thermodynamic
equilibrium, the particle distribution over the states must be
subjected to the principle of detailed balance. We show that this
distribution follows from the equation (4.2). Since the
probability $W_1$ no longer depends on time, the right side of
this equation is equal to zero. Moreover, according to the
principle of detailed balance each term should be equal to zero.
Thus the principle of detailed balance in this case is expressed
by $$ P_{\hh 12}\hs (\hh 1-W_1)\hs W_2=P_{\hh 21}\hs (\hh
1-W_2)\hs W_1\hs . \eqno (4.6) $$ We substitute the expression
(4.4) into this equation. After simple transformations we obtain
the equality $$ \frac{\hs 1-W_1\hh}{W_1}\hs e^{\hh
-\hh\beta\hh\overline\varepsilon_1}= \frac{\hs 1-W_2\hh}{W_2}\hs
e^{\hh -\hh\beta\hh\overline\varepsilon_2}\hs , $$ in which the
left side depends on $\alpha_1$, and the right one - on
$\alpha_2$. This is only possible if both sides are equal to the
same constant value. Denote this value $e^{\hh -\hh\beta\hh\mu}$,
where $\mu$ is the chemical potential. We get the equation $$
\frac{\hs 1-W\hh}{W}=e^{\hh\beta\hh
(\hh\overline\varepsilon-\mu\hh)} \hs . \eqno (4.7) $$ Using the
equality (4.5) we could show that this equation has an anisotropic
solution [15].
\par
From the equation (4.7) we find that the equilibrium distribution
of noninteracting particles over states is described by the Fermi
- Dirac function $$ W=\frac{1}{\hs 1+e^{\hh\beta\hh (\hh
\varepsilon-\mu\hh)}\hh}\hs . \eqno (4.8) $$

\vskip 7mm \hskip-13pt{\bf 5. The electrons in a metal } \vskip
3mm

Consider electrons in metal. Equation (4.7) is transformed to $$
\ln\hs\frac{\hh 1-w_{\hh\fbk}\hh}{w_{\hh\fbk}}= \beta\hs
(\hh\overline\varepsilon_{\fbk}-\mu\hh )\hs . \eqno (5.1) $$ Here,
the function ${w_{\hh\fbk}}$ describes the electron distribution
over wave vectors $\bf k$, $$
\overline\varepsilon_{\fbk}=\varepsilon_{\fbk} +
\sum\limits_{\fbk^\pr}\hs\varepsilon_{\fbk\fbk^\pr}\hs
w_{\hh\fbk^\pr} \eqno (5.2) $$ is the average energy of an electron,
$\varepsilon_{\fbk}$ is the  energy of an electron without taking
into account its interaction with other electrons,
$\varepsilon_{\fbk\fbk^\pr}$ is the interaction energy of electrons
with wave vectors $\bf k$ and $\bf k^\pr$. \par Assume the following
approximate formula for the interaction energy $$
\varepsilon_{\fbk\fbk^\pr}=I\hs\delta_{\hh\fbk+\fbk^\pr}-
J\hs\delta_{\hh\fbk-\fbk^\pr}\hs , \eqno (5.3) $$ where $I$ is the
repulsion energy of electrons with vectors $\bf k$ and $-\hh\bf
k^\pr$, $J$ is the attraction energy of the electrons with vectors
$\bf k$ and $\bf k^\pr$. The substitution of (5.3) into (5.2) gives
$$ \overline\varepsilon_{\fbk}=\varepsilon_{\fbk}+ I\hs w_{\hh
-\hh\fbk}- J\hs w_{\hh\fbk}\hs , \eqno (5.4) $$ and the equation
(5.1) becomes $$ \ln\hs\frac{\hh
1-w_{\hh\fbk}\hh}{w_{\hh\fbk}}=\beta\hs\bigl(\hh
\varepsilon_{\fbk}+I\hs w_{\hh -\fbk}-J\hs
w_{\hh\fbk}-\mu\hh\bigr)\hs . \eqno (5.5) $$ Replace in this
equation the vector $\bf k$ by the vector $-\hh\bf k$: $$
\ln\hs\frac{\hh 1-w_{\hh -\fbk}\hh}{w_{\hh -\fbk}}=\beta\hs\bigl(\hh
\varepsilon_{\fbk} + I\hs w_{\hh\fbk}-J\hs w_{\hh
-\fbk}-\mu\hh\bigr) \hs , \eqno (5.6) $$ where

$$ \varepsilon_{\hh -\hh\fbk}=\varepsilon_{\fbk}\hs . $$ The
equations (5.5) and (5.6) form a system with two unknowns $w_{\fbk}$
and $w_{\hh -\fbk}$. \par This system has the solutions of two
types. One of them describes the isotropic distribution of electrons
over wave vectors and the other describes the anisotropic
distribution. If $w_{\hh -\hh\fbk}=w_{\hh\fbk}$, then each of the
equations (5.5) and (5.6) turns into the equation $$ \ln\hs\frac{\hh
1-w_{\hh\fbk}\hh}{w_{\hh\fbk}}=\beta\hs\Bigl(\hh
\varepsilon_{\fbk}+(\hh I-J\hh)\hs w_{\hh\fbk}-\mu\hh\Bigr)\hs .
\eqno (5.7) $$ \par In the equations (5.5) and (5.6) the unknown
functions $w_{\hh\fbk}$ and $w_{\hh -\hh\fbk}$ are complex
functions, in which the role of an intermediate variable is played
by the kinetic energy $\varepsilon_{\hh\fbk}$ of an electron:
$w_{\hh -\hh\fbk}=w_{\hh 1}(\varepsilon_{\hh\fbk})$ and
$w_{\hh\fbk}=w_{\hh 2}(\varepsilon_{\hh\fbk})$. The functions
$w_{\hh 1}=w_{\hh 1}(\varepsilon)$ and $w_{\hh 2}=w_{\hh
2}(\varepsilon)$ are the solutions of the equations $$
\ln\hh\frac{\hs 1-w_{\hh 1}\hh}{w_{\hh 1}}=\frac{\hs 2\hs}{\tau}
\hs\Bigl(\hh 2\hs\epsilon +(\hh 1-f\hh)\hs w_{\hh 2}- (\hh
1+f\hh)\hs w_{\hh 1}\hh\Bigr)\hs , \eqno (5.8) $$ $$ \ln\hh\frac{\hs
1-w_{\hh 2}\hh}{w_{\hh 2}}=\frac{\hs 2\hs}{\tau} \hs\Bigl(\hh
2\hs\epsilon +(\hh 1-f\hh)\hs w_{\hh 1}- (\hh 1+f\hh)\hs w_{\hh
2}\hh\Bigr)\hs , \eqno (5.9) $$ where  $$ \epsilon
=\frac{\hs\varepsilon -\mu\hs}{J+I}\hs , \hskip 7mm \tau =\frac{\hs
4\hs\theta\hs}{J+I}\hs , $$ the ratio of the energy $I$ and $J$ is
determined by the parameter $$ f=\frac{\hs J-I\hs}{J+I}\hs . $$
Without loss of generality, we may accept the condition $$ w_{\hh
1}(\epsilon)\leq w_{\hh 2}(\epsilon)\hs . \eqno (5.10) $$ \par Let
us introduce new variables $d$ and $s$ by means of relations $$
w_2-w_1=d\hs , \hskip 2cm w_1+w_2=1+s\hs . \eqno (5.11) $$ Due to
(5.10) the quantity $d$ is non-negative: $d\geq 0$. The maximal
value of $d$ is equal to one: $d\in[\hh 0,\hh 1\hh]$. The quantity
$s$ takes on values from $-\hh 1$ to 1: $s\in [\hh -1,\hh 1\hh]$.
Let us solve equations (5.11) with respect to $w_{\hh 1}$ and
$w_{\hh 2}$: $$ w_1=\frac{1}{\hh 2\hh}\hs (\hh 1+s-d\hh)\hs , \hskip
2cm
   w_2=\frac{1}{\hh 2\hh}\hs (\hh 1+s+d\hh)\hs . \eqno (5.12) $$

\par
We transform equations (5.8) and (5.9) using (5.12) to a form
convenient for their numerical solution. To do this we first
subtract one equation from another, then we add these equations.
As a result, we obtain the following system: $$
\left.\begin{array}{l} \ds\frac{\hs (\hh 1+d\hh )^2-s^{\hh
2}\hh}{\hs (\hh 1-d\hh )^2-s^{\hh 2}\hh}= e^{\hh 4\hh d/\tau}\hs ,
\medskip \\ \ds \epsilon=\frac{\tau}{\hh 8\hh}\hs\ln\hh\frac{\hs
(\hh 1-s\hh )^2-d^{\hh 2}\hh} {\hs (\hh 1+s\hh )^2-d^{\hh
2}\hh}+\frac{1}{\hh 2\hh}\hs (\hh 1+s\hh)\hs f \hs . \\
\end{array}\right\} \eqno (5.13) $$ These equations allow us to
represent the energy $\epsilon$ and the probabilities $w_{\hh 1}$
and $w_{\hh 2}$ as functions of the parameter $d$. Using these
functions it is not difficult to construct the graphs of the
functions $w_{\hh 1}=w_{\hh 1}(\epsilon)$ and $w_{\hh 2}=w_{\hh
2}(\epsilon)$ for different values of temperature. These graphs
are shown in Fig. 1 for the case when the interaction energy of
electrons $I$ and $J$ is such that $J=3\hh I$. At the same time
$f=1/2$.

\par
The solution of equation (5.7) is also a function of electron
kinetic energy $\varepsilon_{\mathcal{}\hh\fbk}$:
$w_{\hh\fbk}=w_{\hh 0}(\varepsilon_{\hh\fbk})$. The function
$w_{\hh 0}=w_{\hh 0}(\epsilon)$ can be found from the equation $$
\ln\hh\frac{\hs 1-w_{\hh 0}\hh}{w_{\hh 0}}=\frac{\hs 4\hs}{\tau}
\hs\bigl(\hh \epsilon -f\hs w_{\hh 0}\hh\bigr)\hs , \eqno (5.14)
$$ which is a corollary of equation (5.7). The solution $w_{\hh
TT0}=w_{\hh 0}(\epsilon)$ of equation (5.14) is among the
solutions of equations (5.8) and (5.9). Indeed, if we put in these
equations $w_{\hh 1}=w_{\hh 2}$, then each of them takes the form
(5.14).
\par
Fig. 1 shows the graphs of all three functions $w_{\hh 0}=w_{\hh
0}(\epsilon)$, $w_{\hh 1}=w_{\hh 1}(\epsilon)$ and $w_{\hh
2}=w_{\hh 2}(\epsilon)$. From this figure it is clear that at
sufficiently low temperatures the distribution function for some
values of the energy $\epsilon$ can take not one but several
values. The multivaluedness  of functions $w=w(\epsilon)$ suggests
that at the same temperature there can be different equilibrium
macrostates of a system of conduction electrons in metal. These
macrostates differ from each other by the distributions of
electrons in Bloch states. In fact, only the macrostate of
electrons, in which their energy is minimal, is realized, provided
that this state is stable and cannot be destroyed by any external
effects.

\par The graphs in Fig. 1 give an idea of the changes in the
distribution of electrons in states that occur when the metal
temperature changes. When the temperature $\tau\geq 1$, only the
isotropic distribution of electrons over wave vectors described by
the function $w_{\hh\fbk}=w_{\hh 0}(\varepsilon_{\hh\fbk})$ is
possible. The graph of the function $w_{\hh 0}=w_{\hh
0}(\varepsilon)$ for all values of the temperature passes though
the point $\Omega$ with coordinates $\varepsilon =\mu +\frac12\hh
(J-I)$ and $w=\frac12$. As the temperature decreases, the slope of
the curve at this point increases. When the temperature is
sufficiently low, the curve of $w_{\hh 0}=w_{\hh 0}(\varepsilon)$
bends so that it becomes similar to the letter $Z$.

\par For $\tau=1$ on the curve $w_{\hh 0}=w_{\hh 0}(\varepsilon)$
at the point $\Omega$ a closed curve $Z$ originates. Its size
increases as the temperature decreases. The shape of the curve
also changes. The value $\tau=1$ corresponds to the critical
temperature
$$ T_c=\frac{\hs I+J\hs}{4\hs k_B}\hs . \eqno (5.15) $$
At $\tau\in (\tau^{\hh\pr},\hh 1\hh)$, where $\tau^{\hh\pr}$ is
some critical value, the vertical line intersects the curve $Z$ no
more than in two points (the curve {\it 2} in Fig. 1{\it c}). At
$\tau<\tau^{\hh\pr}$ the curve $Z$ bends so that the vertical line
intersects it four times in some places (Fig. 1{\it b}). In the
limit, $\tau\to 0$, the curve $Z$ transforms into a polygon
$AB_1C_1OC_2B_2A$, that looks like the letter $Z$. This polygon is
shown in Fig. 1{\it а}. The curve $w_{\hh 0}=w_{\hh
0}(\varepsilon)$ divides the curve $Z$ into two parts. The lower
part corresponds to $w_{\hh 1}=w_{\hh 1}(\varepsilon)$ and the
upper - to  $w_{\hh 2}=w_{\hh 2}(\varepsilon)$.

\par Let us give more detailed consideration to the distribution
of electrons over states at $T=0$. In the limit, $\tau\to 0$, the
solution of (5.14) takes on the form
$$ w_{\hh 0}(\varepsilon)=\left\{\begin{array}{ccl} 1 & \hbox{if}
& \varepsilon\leq\mu +J-I\hs , \medskip \\ \ds\frac{\varepsilon
-\mu}{\hh J-I\hh} & \hbox{if} & \mu\leq\varepsilon\leq\mu +J-I\hs
,
\medskip \\ 0 & \hbox{if} & \varepsilon\geq\mu\hs .
\end{array}\right. \eqno (5.16) $$ The graph of this function is
shown in Fig. 1{\it a} as the dotted broken line $B_2AOC_1$.

\par In the limit, $\tau\to 0$, the equations (5.8) and (5.9) give
the following dependence of the probability $w$ on the electron
kinetic energy $\varepsilon$, which describes the anisotropic
distribution of electrons over wave vectors:
$$ w(\varepsilon)=\left\{\begin{array}{ccl} 1 & \hbox{if} &
\varepsilon\leq\mu -I\hs , \medskip \\ w_{\hh i}(\varepsilon) &
\hbox{if} & \mu -I\leq\varepsilon\leq\mu +J\hs , \medskip \\ 0 &
\hbox{if} & \varepsilon\geq\mu +J\hs , \end{array}\right. \eqno
(5.17) $$ where $i=1$ or 2. The values of the functions $w_{\hh
1}=w_{\hh 1}(\varepsilon)$ and $w_{\hh 2}=w_{\hh 2}(\varepsilon)$
form pairs, such that $$ w_{\hh 1}(\varepsilon)=0 \hskip 4mm
\hbox{and} \hskip 4mm
   w_{\hh 2}(\varepsilon)=1 \eqno (5.18) $$
at $\mu -I\leq\varepsilon\leq\mu +J$, or $$ w_{\hh
1}(\varepsilon)=\frac{1}{\hh J\hh}\hh(\hh\varepsilon -\mu +I\hh)
\hskip 4mm \hbox{and} \hskip 4mm w_{\hh 2}(\varepsilon)=1 \eqno
(5.19)
$$ at $\mu -I\leq\varepsilon\leq\mu +J-I$, or $$ w_{\hh
1}(\varepsilon)=0 \hskip 4mm \hbox{and} \hskip 4mm
   w_{\hh 2}(\varepsilon)=\frac{1}{\hh J\hh}\hh(\hh\varepsilon -\mu\hh)
\eqno (5.20) $$ at $0\leq\varepsilon\leq\mu +J$. The broken line
$AB_1C_1O$ in Fig. 1{\it а} corresponds to $w_{\hh 1}=w_{\hh
1}(\varepsilon)$, and the line $AB_2C_2O$ -- to $w_{\hh 2}=w_{\hh
2}(\varepsilon)$.

\vskip 7mm \hskip-13pt{\bf 6. Conclusions} \vskip 3mm The quantum
Markov equation for $N$-particle density matrix describes the
evolution of a strictly open system with allowance for the
dissipative operator. From this equation, we obtained the
approximate kinetic equation for a single-particle density matrix.
This equation describes the evolution of a system of identical
particles. We obtained the integral equation for the function of
distribution of particles in states where a many-particle system
is in thermodynamic equilibrium. We showed that for equilibrium
systems of interacting particles the distribution over wave
vectors is anisotropic.

\newpage

\vskip 7mm \hskip-13pt{\bf References} \vskip 3mm

\hskip-13pt[1] J. von Neumann, Mathematical Foundations of Quantum
Mechanics (Nauka, Moscow, 1964). \vskip 1mm

\hskip-13pt[2] Y.R.Shen, Phys. Rev. 155 (1967) 921. \vskip 1mm

\hskip-13pt[3] M.Grover, R.Silbey, Chem. Phys. 52 (1970) 2099, 54
(1971) 4843. \vskip 1mm

\hskip-13pt[4] A.Kossakowski, Rep. Math. Phys. 3 (1972) 247. \vskip
1mm

\hskip-13pt[5] V.Gorini, A.Kossakowski and E.C.G.Sudarshan, J. Math.
Phys. 17 (1976) 821. \vskip 1mm

\hskip-13pt[6] G.Lindblad, Commun. Math. Phys. 48 (1976) 119. \vskip
1mm

\hskip-13pt[7] V.Gorini, A.Frigerio, N.Verri, A.Kossakowski and
E.C.G.Sudarshan, Rep. Math. Phys. 13 (1978) 149. \vskip 1mm

\hskip-13pt[8] K.Blum, Density Matrix Theory and Application
(Plenum, New York, London, 1981). \vskip 1mm

\hskip-13pt[9] R.L.Stratonovich, Nonlinear Nonequilibrium
Thermodynamics (Nauka, Moscow, 1985). \vskip 1mm

\hskip-13pt[10] R.Alicki, K.Lendy, Quantum Dynamical Semigroups and
Applications. Lecture Notes in Physics. \vskip 1mm vol. 286
(Springer-Verlag, Berlin, 1987). \vskip 1mm

\hskip-13pt[11] J.L.Neto, Ann. Phys. (NY) 173 (1987) 443. \vskip 1mm

\hskip-13pt[12] B.V.Bondarev, Physica A 176 (1991) 366. \vskip 1mm

\hskip-13pt[13] B.V.Bondarev, Physica A 183 (1992) 159. \vskip 1mm

\hskip-13pt[14] B.V.Bondarev, Teoret. Mat. Fis. 100 (1994) 33.
\vskip 1mm

\newpage

\setcounter{page}{11} \makeatletter\renewcommand{\@oddhead} { \hfil
---\hs\hs\hs\thepage\hs\hs\hs ---\hskip 2mm\hfil }
\makeatletter\renewcommand{\@oddfoot}{}

\hskip-13pt {\Large\bf Quantum master equation for a system of
identical particles } \vskip 2mm

\hskip-13pt {\Large\bf and the anisotropic distribution
 of the interacting electrons } \vskip 5mm

\hskip-13pt Boris V. Bondarev \vskip 5mm

\vskip -8mm \unitlength=1mm
\centerline{\begin{picture}(83,77)\put(-2,59){\it а{\rm )}}
\put(-1,52.5){$B_2$}\put(61,47){$A$}\put(78,52.5){$C_2$}
\put(0.5,13.5){$B_1$}\put(16.5,11){$O$}\put(81,12){$C_1$}\put(77,12){$J$}
\put(-2,9){\vector(1,0){87}}\put(77,4){$\varepsilon-\mu$}
\put(20,9){\vector(0,1){51}}\put(22,58.5){$w$}\put(17,51.5){1}
\multiput(0,9.5)(0,1.99){21}{\line(0,1){1.4}}
\put(0,9){\line(0,-1){1}}\put(-2,4){$-\hh I$}
\put(20,9){\line(0,-1){1}}\put(19.3,4){$0$}
\multiput(60,10)(0,1.98){20}{\line(0,1){1.4}}
\put(60,9){\line(0,-1){1}}\put(55.9,4){$J-I$}
\multiput(80,9.5)(0,1.98){21}{\line(0,1){1.4}}
\put(40,30){\circle*{0.7}}\put(36.5,29.5){$\Omega$}
\put(31,33){$Z$}\put(45,24){$Z$}
\multiput(-2,49.5)(2.02,0){31}{\line(1,0){1.4}}
\put(20,10.5){\multiput(0,0)(1.39,1.35){29}
{\unitlength=1mm\special{em:linewidth 0.3pt}
\put(0,0){\special{em:moveto}}\put(1,1){\special{em:lineto}} }}
\multiput(20,10.5)(2.02,0){32}{\line(1,0){1.4}}
\put(0,50){\line(1,0){60}} \put(0,50.5){\line(1,0){80}}
\put(0,9.5){\line(1,0){80}} \put(20,10){\line(1,0){60}}
\put(0,9.5){\unitlength=1mm\special{em:linewidth 0.3pt}
\put(0,0){\special{em:moveto}} \put(60,40.5){\special{em:lineto}} }
\put(20,10){\unitlength=1mm\special{em:linewidth 0.3pt}
\put(0,0){\special{em:moveto}} \put(60,40.5){\special{em:lineto}} }
\end{picture}}

\vskip 9mm \unitlength=1mm
\centerline{\begin{picture}(82,57)\put(-2,59){\it b\hs{\rm )}}
\put(-2,10){\vector(1,0){87}}\put(77,5){$\varepsilon-\mu$}
\put(20,10){\vector(0,1){51}}\put(22,59.5){$w$}\put(17,51.5){1}
\put(0,10){\line(0,-1){1}}\put(-2,5){$-\hh I$}
\put(80,10){\line(0,-1){1}}\put(79,12){$J$}
\put(20,10){\line(0,-1){1}}\put(19.3,5){$0$}
\multiput(60,11)(0,1.98){20}{\line(0,1){1.4}}
\put(60,10){\line(0,-1){1}}\put(55.9,5){$J-I$}
\multiput(-2,50)(2.02,0){31}{\line(1,0){1.4}}
\put(40,30){\circle*{0.7}}\put(36.5,29.5){$\Omega$}
\put(29,33){$Z$}\put(48.5,25){$Z$}
\put(20,10){\unitlength=1mm\special{em:linewidth 0.3pt}
\put(20.00,0.9)  {\special{em:moveto}} \put(18.50,1.0)
{\special{em:lineto}} \put(17.00,1.2)  {\special{em:lineto}}
\put(15.02,1.489)  {\special{em:lineto}} \put(12.69,1.990)
{\special{em:lineto}} \put(11.18,2.442)  {\special{em:lineto}}
\put(10.06,2.874)  {\special{em:lineto}} \put(9.200,3.295)
{\special{em:lineto}} \put(8.512,3.710)  {\special{em:lineto}}
\put(7.514,4.528)  {\special{em:lineto}} \put(7.148,4.932)
{\special{em:lineto}} \put(6.850,5.336)  {\special{em:lineto}}
\put(6.610,5.736)  {\special{em:lineto}} \put(6.418,6.136)
{\special{em:lineto}} \put(6.157,6.932)  {\special{em:lineto}}
\put(6.024,7.724)  {\special{em:lineto}} \put(6.093,9.692)
{\special{em:lineto}} \put(6.545,11.64)  {\special{em:lineto}}
\put(7.236,13.58)  {\special{em:lineto}} \put(8.080,15.48)
{\special{em:lineto}} \put(9.016,17.37)  {\special{em:lineto}}
\put(10.00,19.22)  {\special{em:lineto}} \put(11.00,21.04)
{\special{em:lineto}} \put(11.98,22.81)  {\special{em:lineto}}
\put(12.92,24.52)  {\special{em:lineto}} \put(13.79,26.17)
{\special{em:lineto}} \put(14.58,27.74)  {\special{em:lineto}}
\put(15.25,29.23)  {\special{em:lineto}} \put(15.79,30.62)
{\special{em:lineto}} \put(16.18,31.91)  {\special{em:lineto}}
\put(16.43,33.08)  {\special{em:lineto}} \put(16.43,34.00)
{\special{em:lineto}} }
\put(20,10){\unitlength=1mm\special{em:linewidth 0.3pt}
\put(16.43,34.00)  {\special{em:moveto}} \put(16.43,35.08)
{\special{em:lineto}} \put(16.43,35.08)  {\special{em:lineto}}
\put(16.18,35.91)  {\special{em:lineto}} \put(15.79,36.62)
{\special{em:lineto}} \put(15.25,37.23)  {\special{em:lineto}}
\put(14.58,37.74)  {\special{em:lineto}} \put(13.79,38.17)
{\special{em:lineto}} \put(12.92,38.52)  {\special{em:lineto}}
\put(11.98,38.81)  {\special{em:lineto}} \put(11.00,39.04)
{\special{em:lineto}} \put(10.00,39.22)  {\special{em:lineto}}
\put(9.016,39.37)  {\special{em:lineto}} \put(8.080,39.48)
{\special{em:lineto}} \put(7.236,39.58)  {\special{em:lineto}}
\put(6.545,39.64)  {\special{em:lineto}} \put(6.093,39.69)
{\special{em:lineto}} \put(5.993,39.71)  {\special{em:lineto}}
\put(6.157,39.73)  {\special{em:lineto}} \put(6.610,39.74)
{\special{em:lineto}} \put(7.148,39.73)  {\special{em:lineto}}
\put(8.512,39.71)  {\special{em:lineto}} \put(9.200,39.70)
{\special{em:lineto}} \put(10.06,39.67)  {\special{em:lineto}}
\put(11.18,39.64)  {\special{em:lineto}} \put(12.69,39.59)
{\special{em:lineto}} \put(15.02,39.49)  {\special{em:lineto}}
\put(20.00,39.15)  {\special{em:lineto}} }
\put(20,10){\unitlength=1mm\special{em:linewidth 0.3pt}
\put(20.00,39.15)  {\special{em:moveto}} \put(22.50,38.90)
{\special{em:lineto}} \put(24.98,38.51)  {\special{em:lineto}}
\put(27.31,38.01)  {\special{em:lineto}} \put(28.82,37.56)
{\special{em:lineto}} \put(29.94,37.13)  {\special{em:lineto}}
\put(30.80,36.70)  {\special{em:lineto}} \put(31.49,36.29)
{\special{em:lineto}} \put(32.04,35.88)  {\special{em:lineto}}
\put(32.49,35.47)  {\special{em:lineto}} \put(32.86,35.07)
{\special{em:lineto}} \put(33.15,34.66)  {\special{em:lineto}}
\put(33.39,34.26)  {\special{em:lineto}} \put(33.58,33.86)
{\special{em:lineto}} \put(33.84,33.07)  {\special{em:lineto}}
\put(33.98,32.28)  {\special{em:lineto}} \put(34.01,31.49)
{\special{em:lineto}} \put(33.96,30.70)  {\special{em:lineto}}
\put(33.84,29.92)  {\special{em:lineto}} \put(33.67,29.14)
{\special{em:lineto}} \put(33.46,28.36)  {\special{em:lineto}}
\put(32.77,26.42)  {\special{em:lineto}} \put(31.92,24.52)
{\special{em:lineto}} \put(30.98,22.63)  {\special{em:lineto}}
\put(30.00,20.78)  {\special{em:lineto}} \put(29.00,18.96)
{\special{em:lineto}} \put(28.02,17.19)  {\special{em:lineto}}
\put(27.08,15.48)  {\special{em:lineto}} \put(26.21,13.83)
{\special{em:lineto}} \put(25.42,12.26)  {\special{em:lineto}}
\put(24.75,10.77)  {\special{em:lineto}} \put(24.21,9.380)
{\special{em:lineto}} \put(23.82,8.092)  {\special{em:lineto}}
\put(23.57,6.916)  {\special{em:lineto}} \put(23.57,4.916)
{\special{em:lineto}} }
\put(20,10){\unitlength=1mm\special{em:linewidth 0.3pt}
\put(23.57,4.916)  {\special{em:moveto}} \put(23.82,4.092)
{\special{em:lineto}} \put(24.21,3.378)  {\special{em:lineto}}
\put(24.75,2.769)  {\special{em:lineto}} \put(25.42,2.257)
{\special{em:lineto}} \put(26.21,1.830)  {\special{em:lineto}}
\put(27.08,1.479)  {\special{em:lineto}} \put(28.02,1.192)
{\special{em:lineto}} \put(29.00,0.961)  {\special{em:lineto}}
\put(30.00,0.776)  {\special{em:lineto}} \put(30.98,0.630)
{\special{em:lineto}} \put(31.92,0.514)  {\special{em:lineto}}
\put(32.77,0.425)  {\special{em:lineto}} \put(33.46,0.357)
{\special{em:lineto}} \put(33.90,0.308)  {\special{em:lineto}}
\put(33.98,0.276)  {\special{em:lineto}} \put(33.39,0.264)
{\special{em:lineto}} \put(31.49,0.290)  {\special{em:lineto}}
\put(30.80,0.305)  {\special{em:lineto}} \put(29.94,0.326)
{\special{em:lineto}} \put(28.82,0.358)  {\special{em:lineto}}
\put(27.31,0.410)  {\special{em:lineto}} \put(24.98,0.511)
{\special{em:lineto}} \put(24.00,0.6)  {\special{em:lineto}}
\put(22.00,0.7)  {\special{em:lineto}} \put(20.00,0.9)
{\special{em:lineto}} }
\put(20,10){\unitlength=1mm\special{em:linewidth 0.3pt}
\put(2.364,39.00) {\special{em:moveto}} \put(6.048,38.50)
{\special{em:lineto}} \put(8.552,38.00) {\special{em:lineto}}
\put(10.42,37.50) {\special{em:lineto}} \put(11.88,37.00)
{\special{em:lineto}} \put(13.06,36.50) {\special{em:lineto}}
\put(14.02,36.00) {\special{em:lineto}} \put(15.54,35.00)
{\special{em:lineto}} \put(16.66,34.00) {\special{em:lineto}}
\put(17.50,33.00) {\special{em:lineto}} \put(18.14,32.00)
{\special{em:lineto}} \put(18.63,31.00) {\special{em:lineto}}
\put(19.02,30.00) {\special{em:lineto}} \put(19.53,28.00)
{\special{em:lineto}} \put(19.81,26.00) {\special{em:lineto}}
\put(19.94,24.00) {\special{em:lineto}} \put(19.99,22.00)
{\special{em:lineto}} \put(20.00,20.00) {\special{em:lineto}}
\put(20.01,18.00) {\special{em:lineto}} \put(20.06,16.00)
{\special{em:lineto}} \put(20.19,14.00) {\special{em:lineto}}
\put(20.47,12.00) {\special{em:lineto}} \put(20.98,10.00)
{\special{em:lineto}} \put(21.37,9.00) {\special{em:lineto}}
\put(21.86,8.00) {\special{em:lineto}} \put(22.50,7.00)
{\special{em:lineto}} \put(23.34,6.00) {\special{em:lineto}}
\put(24.46,5.00) {\special{em:lineto}} \put(25.98,4.00)
{\special{em:lineto}} \put(28.12,3.00) {\special{em:lineto}}
\put(29.06,2.667) {\special{em:lineto}} \put(30.15,2.333)
{\special{em:lineto}} \put(31.45,2.00) {\special{em:lineto}}
\put(33.02,1.667) {\special{em:lineto}} \put(35.01,1.333)
{\special{em:lineto}} \put(37.63,1.000) {\special{em:lineto}}
\put(41.44,0.667) {\special{em:lineto}} } \end{picture}}

\vskip 9mm \unitlength=1mm
\centerline{\begin{picture}(82,57)\put(-2,59){\it c\hs{\rm )}}
\put(45.7,42.5){\it 1}\put(42.5,35.3){\it 2} \put(51,12.2){\it
1}\put(53.5,19.5){\it 2}
\put(-2,10){\vector(1,0){87}}\put(77,5){$\varepsilon-\mu$}
\put(20,10){\vector(0,1){51}}\put(22,59.5){$w$}\put(17,51.5){1}
\put(0,10){\line(0,-1){1}}\put(-2,5){$-\hh I$}
\put(20,10){\line(0,-1){1}}\put(19.3,5){$0$}
\multiput(60,11)(0,1.98){20}{\line(0,1){1.4}}
\put(60,10){\line(0,-1){1}}\put(55.9,5){$J-I$}
\multiput(-2,50)(2.02,0){31}{\line(1,0){1.4}}
\put(40,30){\circle*{0.7}}\put(40.2,30.4){$\Omega$}
\multiput(39.5,30)(-2,0){10}{\line(-1,0){1.4}}
\put(20,10){\line(-1,0){1}}\put(17,29){$\frac12$}
\multiput(40,29.5)(0,-2){10}{\line(0,-1){1.4}}
\put(40,10){\line(0,-1){1}}\put(32.3,5){$\frac12(J-I\hh )$}
\put(20,10){\unitlength=1mm\special{em:linewidth 0.3pt}
\put(20.00,4.500)  {\special{em:moveto}} \put(18.01,5.332)
{\special{em:lineto}} \put(16.73,6.044)  {\special{em:lineto}}
\put(15.90,6.616)  {\special{em:lineto}} \put(14.74,7.604)
{\special{em:lineto}} \put(13.95,8.500)  {\special{em:lineto}}
\put(13.64,8.924)  {\special{em:lineto}} \put(13.14,9.752)
{\special{em:lineto}} \put(12.60,10.94)  {\special{em:lineto}}
\put(12.10,12.82)  {\special{em:lineto}} \put(11.90,14.62)
{\special{em:lineto}} \put(11.89,16.34)  {\special{em:lineto}}
\put(12.00,18.01)  {\special{em:lineto}} \put(12.19,19.62)
{\special{em:lineto}} \put(12.42,21.16)  {\special{em:lineto}}
\put(12.66,22.64)  {\special{em:lineto}} \put(12.90,24.04)
{\special{em:lineto}} \put(13.10,25.38)  {\special{em:lineto}}
\put(13.28,26.65)  {\special{em:lineto}} \put(13.41,27.84)
{\special{em:lineto}} \put(13.50,28.96)  {\special{em:lineto}}
\put(13.50,30.96)  {\special{em:lineto}} }
\put(20,10){\unitlength=1mm\special{em:linewidth 0.3pt}
\put(13.50,30.96)  {\special{em:moveto}} \put(13.50,30.96)
{\special{em:lineto}} \put(13.41,31.84)  {\special{em:lineto}}
\put(13.28,32.65)  {\special{em:lineto}} \put(13.10,33.38)
{\special{em:lineto}} \put(12.90,34.04)  {\special{em:lineto}}
\put(12.66,34.64)  {\special{em:lineto}} \put(12.42,35.16)
{\special{em:lineto}} \put(12.19,35.62)  {\special{em:lineto}}
\put(12.00,36.01)  {\special{em:lineto}} \put(11.89,36.34)
{\special{em:lineto}} \put(11.90,36.62)  {\special{em:lineto}}
\put(12.10,36.82)  {\special{em:lineto}} \put(12.60,36.94)
{\special{em:lineto}} \put(13.64,36.92)  {\special{em:lineto}}
\put(13.95,36.90)  {\special{em:lineto}} \put(14.74,36.80)
{\special{em:lineto}} \put(15.26,36.73)  {\special{em:lineto}}
\put(15.90,36.62)  {\special{em:lineto}} \put(16.73,36.44)
{\special{em:lineto}} \put(18.01,36.13)  {\special{em:lineto}}
\put(20.00,35.50)  {\special{em:lineto}} }
\put(20,10){\unitlength=1mm\special{em:linewidth 0.3pt}
\put(20.00,35.50)  {\special{em:moveto}} \put(21.99,34.67)
{\special{em:lineto}} \put(23.27,33.96)  {\special{em:lineto}}
\put(24.10,33.38)  {\special{em:lineto}} \put(24.74,32.87)
{\special{em:lineto}} \put(25.26,32.40)  {\special{em:lineto}}
\put(25.69,31.94)  {\special{em:lineto}} \put(26.36,31.08)
{\special{em:lineto}} \put(27.40,29.06)  {\special{em:lineto}}
\put(27.90,27.18)  {\special{em:lineto}} \put(28.10,25.38)
{\special{em:lineto}} \put(28.11,23.66)  {\special{em:lineto}}
\put(28.00,21.99)  {\special{em:lineto}} \put(27.81,20.38)
{\special{em:lineto}} \put(27.58,18.84)  {\special{em:lineto}}
\put(27.34,17.36)  {\special{em:lineto}} \put(27.10,15.96)
{\special{em:lineto}} \put(26.90,14.62)  {\special{em:lineto}}
\put(26.72,13.35)  {\special{em:lineto}} \put(26.59,12.16)
{\special{em:lineto}} \put(26.50,11.04)  {\special{em:lineto}}
\put(26.50,9.040)  {\special{em:lineto}} }
\put(20,10){\unitlength=1mm\special{em:linewidth 0.3pt}
\put(26.50,9.040)  {\special{em:moveto}} \put(26.59,8.156)
{\special{em:lineto}} \put(26.72,7.348)  {\special{em:lineto}}
\put(26.90,6.616)  {\special{em:lineto}} \put(27.10,5.956)
{\special{em:lineto}} \put(27.34,5.364)  {\special{em:lineto}}
\put(27.58,4.840)  {\special{em:lineto}} \put(27.81,4.384)
{\special{em:lineto}} \put(28.00,3.988)  {\special{em:lineto}}
\put(28.11,3.655)  {\special{em:lineto}} \put(28.10,3.384)
{\special{em:lineto}} \put(27.90,3.183)  {\special{em:lineto}}
\put(27.40,3.065)  {\special{em:lineto}} \put(26.36,3.074)
{\special{em:lineto}} \put(24.10,3.384)  {\special{em:lineto}}
\put(23.27,3.554)  {\special{em:lineto}} \put(21.99,3.866)
{\special{em:lineto}} \put(20.00,4.500)  {\special{em:lineto}} }
\put(20,10){\unitlength=1mm\special{em:linewidth 0.3pt}
\put(-15.95,39.00) {\special{em:moveto}} \put(-10.18,38.50)
{\special{em:lineto}} \put(-6.166,38.00)  {\special{em:lineto}}
\put(-3.121,37.50)  {\special{em:lineto}} \put(-0.685,37.00)
{\special{em:lineto}} \put(1.332,36.50)  {\special{em:lineto}}
\put(3.042,36.00)  {\special{em:lineto}} \put(5.811,35.00)
{\special{em:lineto}} \put(7.981,34.00)  {\special{em:lineto}}
\put(9.744,33.00)  {\special{em:lineto}} \put(11.21,32.00)
{\special{em:lineto}} \put(12.45,31.00)  {\special{em:lineto}}
\put(13.52,30.00)  {\special{em:lineto}} \put(15.29,28.00)
{\special{em:lineto}} \put(16.71,26.00)  {\special{em:lineto}}
\put(17.92,24.00)  {\special{em:lineto}} \put(18.99,22.00)
{\special{em:lineto}} \put(20.00,20.00)  {\special{em:lineto}}
\put(21.01,18.00)  {\special{em:lineto}} \put(22.08,16.00)
{\special{em:lineto}} \put(23.29,14.00)  {\special{em:lineto}}
\put(24.71,12.00)  {\special{em:lineto}} \put(26.48,10.00)
{\special{em:lineto}} \put(27.55,9.00)  {\special{em:lineto}}
\put(28.79,8.00)  {\special{em:lineto}} \put(30.26,7.00)
{\special{em:lineto}} \put(32.02,6.00)  {\special{em:lineto}}
\put(34.19,5.00)  {\special{em:lineto}} \put(36.96,4.00)
{\special{em:lineto}} \put(40.69,3.00)  {\special{em:lineto}}
\put(46.17,2.00)  {\special{em:lineto}} \put(55.95,1.00)
{\special{em:lineto}} \put(61.83,0.667)  {\special{em:lineto}} }
\put(20,10){\unitlength=1mm\special{em:linewidth 0.3pt}
\put(20.00,12.50)  {\special{em:moveto}} \put(19.23,13.13)
{\special{em:lineto}} \put(18.11,14.61)  {\special{em:lineto}}
\put(17.53,15.67)  {\special{em:lineto}} \put(17.12,16.60)
{\special{em:lineto}} \put(16.82,17.45)  {\special{em:lineto}}
\put(16.58,18.25)  {\special{em:lineto}} \put(16.40,19.00)
{\special{em:lineto}} \put(16.26,19.72)  {\special{em:lineto}}
\put(16.14,20.40)  {\special{em:lineto}} \put(16.05,21.06)
{\special{em:lineto}} \put(15.98,21.68)  {\special{em:lineto}}
\put(15.92,22.29)  {\special{em:lineto}} \put(15.88,22.87)
{\special{em:lineto}} \put(15.85,23.43)  {\special{em:lineto}}
\put(15.83,23.96)  {\special{em:lineto}} \put(15.83,24.96)
{\special{em:lineto}} }
\put(20,10){\unitlength=1mm\special{em:linewidth 0.3pt}
\put(15.83,24.96)  {\special{em:moveto}} \put(15.85,25.43)
{\special{em:lineto}} \put(15.88,25.87)  {\special{em:lineto}}
\put(15.92,26.29)  {\special{em:lineto}} \put(15.98,26.68)
{\special{em:lineto}} \put(16.05,27.06)  {\special{em:lineto}}
\put(16.14,27.40)  {\special{em:lineto}} \put(16.26,27.72)
{\special{em:lineto}} \put(16.40,28.00)  {\special{em:lineto}}
\put(16.58,28.25)  {\special{em:lineto}} \put(16.82,28.45)
{\special{em:lineto}} \put(17.12,28.60)  {\special{em:lineto}}
\put(17.53,28.67)  {\special{em:lineto}} \put(18.11,28.61)
{\special{em:lineto}} \put(19.23,28.13)  {\special{em:lineto}}
\put(20.00,27.50)  {\special{em:lineto}} }
\put(20,10){\unitlength=1mm\special{em:linewidth 0.3pt}
\put(20.00,27.50)  {\special{em:moveto}} \put(20.77,26.87)
{\special{em:lineto}} \put(21.89,25.39)  {\special{em:lineto}}
\put(22.47,24.33)  {\special{em:lineto}} \put(22.88,23.40)
{\special{em:lineto}} \put(23.18,22.55)  {\special{em:lineto}}
\put(23.42,21.75)  {\special{em:lineto}} \put(23.60,21.00)
{\special{em:lineto}} \put(23.74,20.28)  {\special{em:lineto}}
\put(23.86,19.60)  {\special{em:lineto}} \put(23.95,18.94)
{\special{em:lineto}} \put(24.02,18.32)  {\special{em:lineto}}
\put(24.08,17.71)  {\special{em:lineto}} \put(24.12,17.13)
{\special{em:lineto}} \put(24.15,16.57)  {\special{em:lineto}}
\put(24.17,16.04)  {\special{em:lineto}} \put(24.17,15.04)
{\special{em:lineto}} }
\put(20,10){\unitlength=1mm\special{em:linewidth 0.3pt}
\put(24.17,15.04)  {\special{em:moveto}} \put(24.15,14.57)
{\special{em:lineto}} \put(24.12,14.13)  {\special{em:lineto}}
\put(24.08,13.71)  {\special{em:lineto}} \put(24.02,13.32)
{\special{em:lineto}} \put(23.95,12.94)  {\special{em:lineto}}
\put(23.86,12.60)  {\special{em:lineto}} \put(23.74,12.28)
{\special{em:lineto}} \put(23.60,12.00)  {\special{em:lineto}}
\put(23.42,11.75)  {\special{em:lineto}} \put(23.18,11.55)
{\special{em:lineto}} \put(22.88,11.40)  {\special{em:lineto}}
\put(22.47,11.33)  {\special{em:lineto}} \put(21.89,11.39)
{\special{em:lineto}} \put(20.77,11.87)  {\special{em:lineto}}
\put(20.00,12.50)  {\special{em:lineto}} }
\put(20,10){\unitlength=1mm\special{em:linewidth 0.3pt}
\put(-22.00,38.20) {\special{em:moveto}} \put(-17.94,38.00)
{\special{em:lineto}} \put(-13.95,37.50) {\special{em:lineto}}
\put(-10.74,37.00) {\special{em:lineto}} \put(-5.747,36.00)
{\special{em:lineto}} \put(-1.972,35.00) {\special{em:lineto}}
\put(1.042,34.00) {\special{em:lineto}} \put(5.661,32.00)
{\special{em:lineto}} \put(9.128,30.00) {\special{em:lineto}}
\put(11.90,28.00) {\special{em:lineto}} \put(14.24,26.00)
{\special{em:lineto}} \put(16.30,24.00) {\special{em:lineto}}
\put(18.18,22.00) {\special{em:lineto}} \put(20.00,20.00)
{\special{em:lineto}} \put(21.82,18.00) {\special{em:lineto}}
\put(23.70,16.00) {\special{em:lineto}} \put(25.76,14.00)
{\special{em:lineto}} \put(28.10,12.00) {\special{em:lineto}}
\put(30.87,10.00) {\special{em:lineto}} \put(32.50,9.000)
{\special{em:lineto}} \put(34.34,8.000) {\special{em:lineto}}
\put(36.46,7.000) {\special{em:lineto}} \put(38.96,6.000)
{\special{em:lineto}} \put(41.98,5.000) {\special{em:lineto}}
\put(43.13,4.667) {\special{em:lineto}} \put(44.38,4.333)
{\special{em:lineto}} \put(45.74,4.000) {\special{em:lineto}}
\put(47.24,3.667) {\special{em:lineto}} \put(48.90,3.333)
{\special{em:lineto}} \put(50.74,3.000) {\special{em:lineto}}
\put(52.81,2.667) {\special{em:lineto}} \put(55.18,2.333)
{\special{em:lineto}} \put(57.94,2.000) {\special{em:lineto}}
\put(61.24,1.667) {\special{em:lineto}} } \end{picture}}

\newpage

\setcounter{page}{12} \makeatletter\renewcommand{\@oddhead} { \hfil
---\hs\hs\hs\thepage\hs\hs\hs ---\hskip 2mm\hfil }
\makeatletter\renewcommand{\@oddfoot}{}

\hskip-13pt {\Large\bf Quantum master equation for a system of
identical particles } \vskip 2mm

\hskip-13pt {\Large\bf and the anisotropic distribution
 of the interacting electrons } \vskip 5mm

\hskip-13pt Boris V. Bondarev \vskip 5mm

{\it Fig. 1. The distribution function of conductivity electrons
in energies for the case when \linebreak J=3\hh I, and at
different temperatures $\tau$: {\it a{\rm )}} $\tau=0$; {\it
b\hs{\rm )}} $\tau=0.5$; {\it c\hs{\rm )}} 1 -- $\tau=0.75$;
\linebreak 2 -- $\tau=0.95$. } \vskip 4pt

\end{document}